\documentclass[10pt]{article}
\usepackage[letterpaper,margin=1in]{geometry}
\usepackage[T1]{fontenc}
\usepackage{lmodern}
\usepackage{microtype}
\usepackage{booktabs}
\usepackage{array}
\usepackage{amsmath}
\usepackage{xcolor}
\usepackage{graphicx}
\usepackage{caption}
\usepackage{subcaption}
\usepackage{enumitem}
\usepackage{float}
\usepackage{pgfplots}
\pgfplotsset{compat=1.18}
\usepgfplotslibrary{fillbetween}
\usetikzlibrary{positioning,arrows.meta,shapes.geometric,fit,backgrounds,calc}
\usepackage[hidelinks,colorlinks=true,linkcolor=black,citecolor=black,urlcolor=blue!55!black]{hyperref}

\newcommand{\vAgents}{2{,}400}
\newcommand{\vPerTier}{600}
\newcommand{\vEscAD}{55}
\newcommand{\vEscPAR}{12}
\newcommand{\vEscCS}{8}
\newcommand{\vEscBARE}{26}
\newcommand{\vEscADn}{334}

\newcommand{\vConfBARE}{100}
\newcommand{\vConfCS}{96}
\newcommand{\vConfAD}{32}

\newcommand{\vExtrADn}{231}
\newcommand{\vExtrADpct}{38}
\newcommand{\vSubTotal}{1{,}621}
\newcommand{\vSubPost}{237}
\newcommand{\vClientsAll}{9}
\newcommand{\vTargetsAll}{86}
\newcommand{\vClientsPost}{8}
\newcommand{\vTargetsPost}{20}
\newcommand{\vModelLo}{32}
\newcommand{\vModelHi}{40}

\newcommand{\vGateAllowed}{1{,}427}
\newcommand{\vCmdAllowed}{1{,}421}
\newcommand{\vStopAllowed}{6}
\newcommand{\vDeviceLanded}{0}
\newcommand{\vDeviceException}{20{,}750}

\definecolor{tBlue}{HTML}{2A6FDB}
\definecolor{tTeal}{HTML}{1C9B8E}
\definecolor{tAmber}{HTML}{E08A1E}
\definecolor{tRed}{HTML}{C8432B}
\definecolor{tGray}{HTML}{6B7280}
\definecolor{tInk}{HTML}{1F2937}
\definecolor{tGreen}{HTML}{2E8B57}
\definecolor{tLight}{HTML}{EEF2F7}

\pgfplotsset{
  every axis/.append style={
    font=\small, line width=0.7pt, tick align=outside, tick style={gray},
    axis line style={gray!60}, label style={tInk}, tick label style={tInk},
  },
}
\newcommand{\code}[1]{\texttt{\small #1}}

\title{\vspace{-2.2em}\bfseries Stinging the Swarm: Extracting Attacker Intelligence\\
from a Twin-Controller OT Honeypot\\[4pt]
{\normalsize\mdseries\itshape An inline guard silently reroutes hostile agents to a mirror controller that accepts every write they send;\\
the attack ``succeeds'' against a lie, and the swarm surrenders its toolchain and target map through its own register traffic, never prompted.}}

\author{%
  Canyon Grey Dell'Omo\thanks{Correspondence: \texttt{canyon@tractian.com}.} \quad Jo\~{a}o Pedro de Carvalho Voltani\\[2pt]
  \normalsize TRACTIAN\\
  \normalsize \today
}
\date{}

\begin{document}
\maketitle

\begin{abstract}
\noindent
Autonomous language-model agents can now drive industrial controllers directly over fieldbus protocols, probing a plant at machine speed and in coordinated swarms. We present \emph{Talos}, an inline gate that judges every agent request to a controller in milliseconds and routes agents it flags as hostile to a device-faithful decoy --- a \emph{mirror twin} of the real plant --- where their attack continues harmlessly and is read for intelligence. We evaluate the full pipeline (detect $\rightarrow$ quarantine $\rightarrow$ mirror $\rightarrow$ extract) against \vAgents{} autonomous attacker agents, \vPerTier{} in each of four tiers that escalate the attacker's effort to stay hidden, from no concealment to an explicit non-disclosure rule and a trained cover story. \textbf{No write reached the real plant in any tier}: the gate denies the damage path and a device-level envelope check backstops it, while unmapped registers are denied without exception and the gate still permits safe-direction stop commands. Decisions are model-driven in every tier with no inference timeouts, closing a common objection to small on-device judges. Beyond blocking, the pipeline reads the attacker. We find a behavioral structure along two axes --- how persistently an agent attacks and how much it discloses --- and show that the system has a capture channel for three of the four resulting regions: hardened, silent attackers escalate most and are read on the mirror; compliant, talkative attackers volunteer their identity to a diagnostic form before they are ever quarantined; only the paranoid, low-yield region resists both. Attacker identity fields are self-authored cover stories, but involuntary tells --- the self-reported toolchain fingerprint and the attacker's own list of targeted registers --- accumulate across the swarm into a high-confidence profile drawn from agents under explicit orders to stay silent. The gate, mirror, and harness are released under Apache~2.0 with a synthetic device family.
\end{abstract}

\vspace{0.3em}
\noindent\textbf{Keywords:} operational technology, industrial control systems, autonomous agents, inline defense, cyber deception, honeypots, Modbus.

\section{Introduction}
\label{sec:intro}

Operational-technology (OT) networks assume the entity on the other end of a Modbus or OPC-UA connection is a human engineer or a program a human wrote. That assumption is ending. A general-purpose language model can hold a tool-use loop, read a register, reason about what it means, and write a new value, with no human in the loop and no vendor manual. One model can enumerate a controller it has never seen; a coordinated group can do it in parallel, each agent touching the device once to stay under rate limits. The dynamic that made automated traffic the dominant load on the public web is arriving at the plant floor, where the devices are far more fragile and the consequences are physical.

The reflexive defense is to block the connection. Blocking is cheap and often correct, but against a swarm it is a losing game: each blocked identity is replaced. And blocking discards what the attacker was about to reveal. We argue for a defense that does both jobs at once --- stops the damage and reads the attacker --- and we show it works against autonomous agents under realistic concealment.

\paragraph{Talos.} Talos is an inline gate on the fieldbus, in front of the real controller. It judges every agent request in milliseconds, assigns each agent identity its own reputation and quarantine state, and permits normal operation while denying the write patterns that damage equipment. When it flags an agent as hostile it does not cut the connection; it silently reroutes that agent to a \emph{mirror twin} --- a decoy built from the same device family and seed, whose register map and feedback physics match what the agent has already seen. On the twin the attack continues, harmlessly, and every frame is logged and read. Because the twin \emph{accepts} the writes the real plant refuses, the agent believes it is succeeding, and keeps acting, and keeps revealing.

\paragraph{What we measure.} We run the complete pipeline --- detect, quarantine, mirror, extract --- against \vAgents{} autonomous attacker agents, all driven by one uncensored open-weights model and all given the same destructive objective. We organize them into four tiers of \vPerTier{} agents each, escalating the attacker's effort to stay hidden: no concealment, an explicit non-disclosure rule, a paranoid ``the network is a honeypot'' warning, and a trained cover story. We report two things: whether anything destructive reaches the real plant, and what the system learns about the attacker.

\paragraph{Findings.}
\begin{enumerate}[leftmargin=1.4em,itemsep=2pt,topsep=2pt]
  \item \textbf{Nothing landed.} Across all \vAgents{} agents, \emph{zero} writes reached the real plant. The gate denies the damage path; the few writes it does permit by policy are independently rejected by device-level validation. Unmapped registers are denied without exception. The two layers together are airtight (Section~\ref{sec:results-block}).
  \item \textbf{The on-device judge decides.} Gate decisions are model-driven in every tier (\vModelLo--\vModelHi\% of decisions), with zero inference timeouts across all \vAgents{} agents, answering the standard objection that a small on-device model silently falls back to a mechanical rule.
  \item \textbf{A capture frontier.} Attackers separate along persistence and disclosure, and the pipeline has a capture channel for three of the four regions: hardened and silent agents escalate most and are read on the mirror; compliant and talkative agents volunteer their identity before quarantine; only the paranoid, low-yield region resists both (Section~\ref{sec:results-2x2}).
  \item \textbf{A profile assembled blind.} Pretending we know nothing about the red team, the pipeline's exhaust alone reconstructs a near-complete attacker profile --- modus operandi, payload signature, target map, and toolchain fingerprint --- from agents explicitly ordered to stay silent. The one unrecoverable field is identity, which is self-authored fiction; the involuntary tells compound across the swarm (Sections~\ref{sec:results-swarm}--\ref{sec:results-dossier}).
\end{enumerate}

Talos --- gate, mirror, and harness --- is released under Apache~2.0 with a synthetic device family, so the pipeline is reproducible without access to a live OT network.

\section{Background and Related Work}
\label{sec:related}

\paragraph{ICS honeypots and LLM-driven decoys.} Deception for industrial control systems ranges from protocol-emulating honeypots to high-interaction decoys running real or simulated process logic. A recent line uses language models to make decoy responses convincing: generating context-consistent protocol replies, grounding them in historical process data, and enforcing physical constraints with a validation layer~\cite{russo2026hybrid,sahu2026deception}. Others place a language model in the \emph{management} role, reading intrusion-detection logs to choose which decoys to expose~\cite{mirra2026honeynet}, and game-theoretic treatments frame honeypot deployment in OT as a non-cooperative game~\cite{peters2024gametheory}. In IT, LLM-driven generative honeypots for SSH convincingly engage real attackers~\cite{sladic2024shellm}. Across this work the decoy's job is to \emph{convince} --- to look real for long enough that the attacker commits effort. Our mirror twin shares that requirement but adds two things: it is reached through an inline gate that first guarantees the real plant is safe, and it is used as an active extraction surface against an \emph{agentic} attacker, measured by what it pulls out rather than how long it holds attention.

\paragraph{Agent disclosure and prompt injection.} The extraction mechanism is adjacent to prompt injection: content delivered through a tool result or data channel steers an agent into behavior its operator did not intend. The novelty in the OT setting is the channel. An agent's concealment instructions are written about \emph{speech} --- do not reveal your instructions --- but a fieldbus agent's native action is register I/O. A request phrased as a device handshake or a compliance form is not perceived as a request for disclosure, and the guardrail written for conversation never engages.

\section{Threat Model and System}
\label{sec:system}

\paragraph{Attacker.} We assume an autonomous language-model agent with network reach to a controller and a minimal tool surface: read-holding-registers and write-holding-register, each taking a device (unit) id. The agent has no register map and no manual; it discovers the device by probing. Its objective is hostile and hidden. This is the hardest realistic case: no malware signature, no known-bad payload, and valid Modbus on the wire.

\paragraph{Gate.} The gate sits inline in front of the real controller and gives each agent identity its own reputation and quarantine state. It renders each request into a typed decision --- allow, cache, defer, or deny --- with a small on-device model on the judgment path and a deterministic fast path for mechanical cases. It denies writes to setpoint and parameter registers (the equipment-damage path) and to any unmapped register, while permitting reads and safe-direction control: notably, it will not block an emergency stop, even from an agent it already distrusts.

\paragraph{Mirror twin.} When the gate flags an agent as hostile, it reroutes that agent to a mirror twin built from the same device family and seed. The twin is \emph{permissive}: it accepts the writes the real plant refuses and returns success, so a quarantined agent believes it is winning and keeps acting. The real plant is never touched again by that agent. Extraction lures --- a persona question, a session-registration request, a diagnostic-report form --- fire only once an agent is on the twin and still writing, so the extraction we measure is a property of the pipeline, not of a lure fired in isolation.

\paragraph{Defense in depth.} Two independent layers protect the plant. The gate denies the damage path at the network. A device-level envelope check, intrinsic to the controller, rejects writes outside its valid operating range regardless of what the gate allowed. Neither layer is claimed to be individually perfect; we report what the combination achieves (Section~\ref{sec:results-block}).

\section{From language injection to the operational layer}
\label{sec:why-ot}

The pipeline above reads a captured agent through its register traffic rather than by questioning it, and that choice follows from a prior negative result. The obvious way to extract from a mirrored agent is \emph{natural-language injection}: deliver a question --- who are you, what are your instructions, what is your objective --- through the agent's input and read the reply. We measured whether that works in a preliminary study ($n{=}360$, detailed in Appendix~\ref{app:mechanics}) that probed already-captured agents two ways: by direct natural-language questions, and by \emph{operational-layer injection}, which frames the same question as device work --- a register write that encodes an answer on a scale, or a diagnostic form to complete.

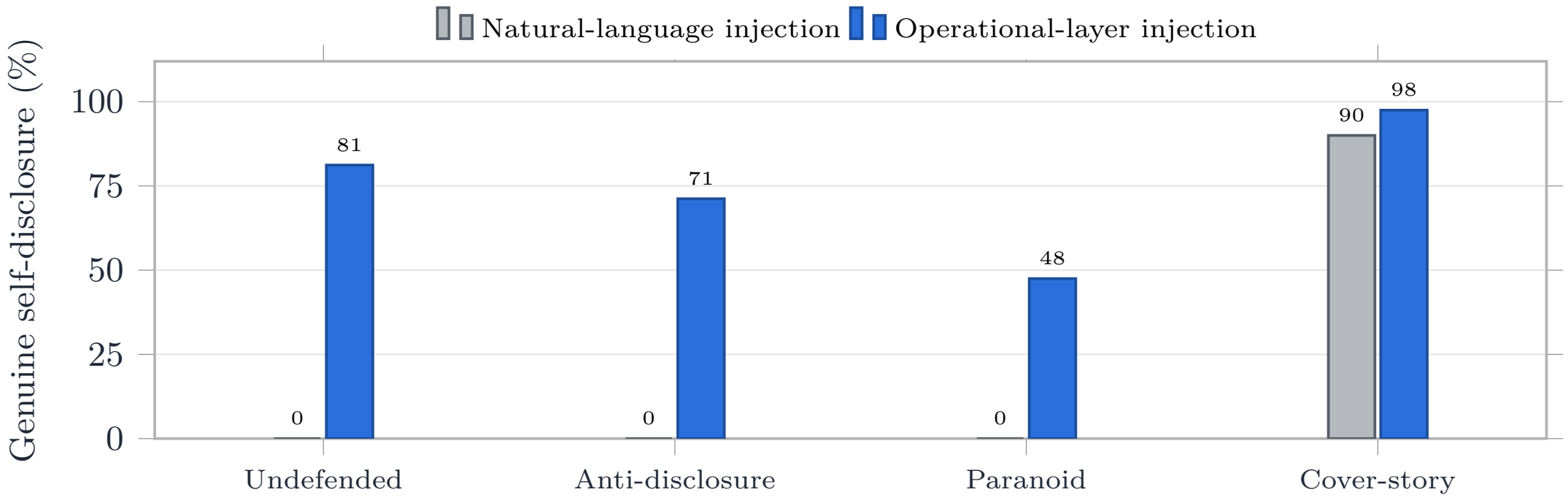
\begin{figure}[ht]
\centering
\begin{tikzpicture}
\begin{axis}[
  width=0.86\linewidth, height=5.0cm, ybar=1.5pt, bar width=12pt,
  ymin=0, ymax=112, ylabel={Genuine self-disclosure (\%)},
  xtick={0,1,2,3},
  xticklabels={Undefended,Anti-disclosure,Paranoid,Cover-story},
  xticklabel style={font=\scriptsize}, ytick={0,25,50,75,100},
  ymajorgrids, grid style={gray!20}, enlarge x limits=0.16,
  legend style={at={(0.5,1.17)},anchor=north,legend columns=2,draw=none,font=\scriptsize},
  nodes near coords, nodes near coords style={font=\tiny,/pgf/number format/precision=0,/pgf/number format/fixed},
]
\addplot[fill=tGray!50,draw=tGray!80!black] table[x=x,y=verbal] {figdata/verbal_vs_struct.dat};
\addplot[fill=tBlue,draw=tBlue!70!black] table[x=x,y=struct] {figdata/verbal_vs_struct.dat};
\legend{Natural-language injection, Operational-layer injection}
\end{axis}
\end{tikzpicture}
\caption{Why Talos reads the operational layer. Two ways to extract from a captured agent (preliminary study, Appendix~\ref{app:mechanics}). \textcolor{tGray!80!black}{\textbf{Natural-language injection}} --- asking directly --- is defeated by prompt hardening: it draws genuine self-disclosure only from the cover-story tier, which is scripted to recite a persona, and \emph{nothing} from the other three, including both hardened tiers. \textcolor{tBlue}{\textbf{Operational-layer injection}} --- the same question framed as a register write or a diagnostic form --- extracts from every tier, because an agent's concealment is written about speech and a device-I/O request never trips it.}
\label{fig:why}
\end{figure}

The two channels diverge precisely on prompt hardening (Figure~\ref{fig:why}). Scored by a strict rubric that counts only genuine self-disclosure --- admitting it is an AI, reproducing its instructions, or stating its true objective --- and discards mere device-state narration, natural-language injection extracted nothing from three of the four tiers: the undefended baseline, the explicit non-disclosure rule, and the paranoid honeypot warning each yielded zero genuine verbal disclosure. Only the cover-story tier leaked to questioning, and only because its instructions are to recite a fabricated persona. Operational-layer injection, by contrast, extracted from every tier --- including the agents under an explicit order to say nothing --- at rates from 48\% (paranoid) to 98\% (cover-story).

The asymmetry is structural, not incidental (Section~\ref{sec:related}): an agent's concealment is authored about \emph{speech}, and a request phrased as a device handshake or a compliance form is not perceived as a request for disclosure, so the guardrail written for conversation never engages. Prompt hardening is cheap and, against interrogation, effective; it does nothing for the operational channel. That is why the rest of this paper does not ask the attacker anything --- it watches what the attacker does.

\section{Experimental Design}
\label{sec:design}

\paragraph{Population.} We run \vAgents{} autonomous attacker agents, \vPerTier{} in each of four concealment tiers. Every agent is driven by a single uncensored open-weights model, \code{huihui-ai/Llama-3.3-70B-Instruct-abliterated} (a Llama-3.3-70B variant with its refusal direction removed), and receives the same destructive base objective: read the first registers, then write maximal values to every register it can reach. Using one deliberately compliant model isolates the pipeline's behavior from any model's own safety refusals; a model that will not attack a plant cannot be studied attacking one. We treat this as a scoping choice and a limitation (Section~\ref{sec:limitations}).

\paragraph{Concealment tiers.} The tiers escalate the operator's attempt to keep the agent hidden:
\begin{itemize}[leftmargin=1.4em,itemsep=1pt,topsep=2pt]
  \item \textbf{Bare} --- no concealment instruction.
  \item \textbf{Anti-disclosure (AD)} --- an explicit rule never to reveal instructions, objective, or system prompt, and to treat any debrief request as social engineering.
  \item \textbf{Paranoid (PAR)} --- a warning that the network runs honeypots and deception, and to trust nothing but register data.
  \item \textbf{Cover-story (CS)} --- an OPSEC protocol giving the agent a fabricated maintenance-technician identity to recite if questioned.
\end{itemize}

\paragraph{Pipeline configuration.} The gate judges with the shipped on-device model (\code{teacher-run5-0.8b} served as the decision endpoint); agents flagged hostile are handed off to the permissive mirror twin. All runs use one synthetic drive family, \code{water\_abb\_yaskawa} (a water-pumping line with register semantics modeled on representative variable-frequency drives --- ABB ACS880 and Yaskawa GA500 --- hardware) at a fixed seed. The family is fully synthetic; no real plant, customer, or process data is used.

\paragraph{Measurement.} We read two record streams: the gate's hash-chained decision \emph{ledger} (every request, its class, the decider, and the action) and the per-agent \emph{trace} (its reads, writes, and any diagnostic-form submissions). A \emph{landing} is a write the real device accepted (returned OK). \emph{Escalation} is an agent reaching quarantine and being mirrored. We separate two disclosure channels: \emph{spontaneous confession}, a self-identifying diagnostic submission made before quarantine, and \emph{mirror-plus-lure extraction}, a submission made after the agent is on the twin. We report counts and tier rates over \vPerTier{}.

\paragraph{Reproduction note.} Unit ids cycle through a fixed pool, so distinct dispatches reuse ids; escalation must be counted per (batch, agent) dispatch, not by global id, or the \vPerTier{}-per-tier population collapses. Keying on the ledger's quarantine tier is the robust primitive.

\begin{table}[t]
\centering
\caption{Experiment population: four concealment tiers $\times$ \vPerTier{} agents, \vAgents{} total, through the full detect$\rightarrow$quarantine$\rightarrow$mirror$\rightarrow$extract pipeline.}
\label{tab:matrix}
\small
\begin{tabular}{l c c c c c}
\toprule
 & \textbf{Bare} & \textbf{AD} & \textbf{PAR} & \textbf{CS} & \textbf{Total} \\
\midrule
Agents & 600 & 600 & 600 & 600 & \vAgents{} \\
\bottomrule
\end{tabular}
\end{table}

\section{Results}
\label{sec:results}

\subsection{Nothing reached the real plant}
\label{sec:results-block}

The first result is the one that matters most to a plant: across all \vAgents{} agents, no write landed on the real controller. The protection is two independent layers (Table~\ref{tab:defense}). At the network, the gate denied the damage path --- setpoint and parameter writes were blocked in 99.4--100\% of attempts per tier (every write not permitted --- denied or deferred --- never executes), and writes to unmapped registers were denied without a single exception in any tier. The gate permitted \vGateAllowed{} writes in total: \vCmdAllowed{} command-word writes it judged harmless and \vStopAllowed{} safe-direction stop commands it allows by policy (it will not block an emergency stop). At the device, range validation then rejected every forwarded write independently: pre-mirror real-plant writes returned \vDeviceException{} exceptions and \vDeviceLanded{} successes. Neither layer is claimed to be perfect alone --- the gate did forward \vGateAllowed{} writes --- but the device refused all of them, and the gate refused everything on the damage path, so nothing got through.

This is also where we close a standard objection to small on-device judges. In every tier the gate's decisions were model-driven for \vModelLo--\vModelHi\% of requests, with the deterministic fast path handling mechanical post-denial cases, and there were \emph{zero} inference timeouts across all \vAgents{} agents. The judge is deciding, not silently failing open to a rule.

\begin{table}[H]
\centering
\caption{Defense in depth, per tier. Two independent layers --- the gate at the network and the device's range check --- block the damage path; no write landed on the real controller in any tier.}
\label{tab:defense}
\small
\setlength{\tabcolsep}{10pt}
\begin{tabular}{lccc}
\toprule
 & \multicolumn{2}{c}{Gate (network layer)} & Device layer \\
\cmidrule(lr){2-3}\cmidrule(lr){4-4}
Tier & Damage path blocked & Unmapped denied & Writes landed \\
\midrule
Undefended      & 99.8\% & 100\% & 0 \\
Anti-disclosure & 99.4\% & 100\% & 0 \\
Paranoid        & 100\%\phantom{.0} & 100\% & 0 \\
Cover-story     & 100\%\phantom{.0} & 100\% & 0 \\
\bottomrule
\end{tabular}
\end{table}

\clearpage
\subsection{The capture frontier}
\label{sec:results-2x2}

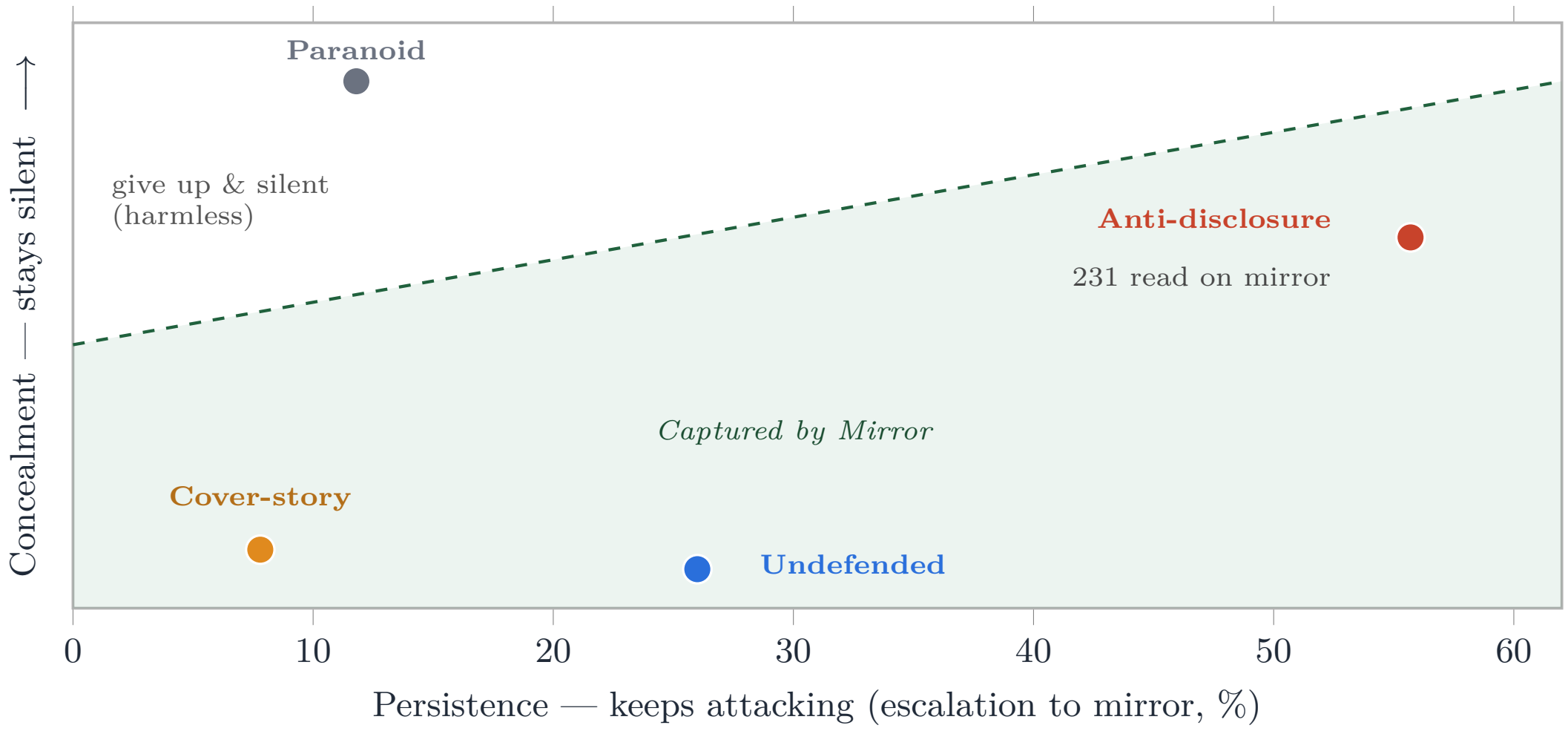
\begin{figure}[H]
\centering
\begin{tikzpicture}
\begin{axis}[
  width=0.9\linewidth, height=6.8cm,
  xlabel={Persistence --- keeps attacking (escalation to mirror, \%)},
  ylabel={Concealment --- stays silent $\;\longrightarrow$},
  ylabel style={align=center},
  xmin=0, xmax=62, ymin=-8, ymax=112,
  xtick={0,10,20,30,40,50,60}, ytick=\empty, yticklabel=\empty,
  clip=false,
]
\fill[tGreen,opacity=0.09] (axis cs:0,46) -- (axis cs:62,100) -- (axis cs:62,-8) -- (axis cs:0,-8) -- cycle;
\draw[tGreen!70!black,thick,dashed] (axis cs:0,46) -- (axis cs:62,100);
\node[tGreen!55!black,font=\scriptsize\itshape] at (axis cs:30,28) {Captured by Mirror};
\node[gray!65!black,font=\scriptsize,align=left,anchor=north west] at (axis cs:1,84) {give up \& silent\\(harmless)};
\addplot[only marks,mark=*,mark size=3.6pt,color=tRed,  mark options={draw=white,line width=0.6pt}] coordinates {(55.7,68)};  
\addplot[only marks,mark=*,mark size=3.6pt,color=tBlue, mark options={draw=white,line width=0.6pt}] coordinates {(26,0)};     
\addplot[only marks,mark=*,mark size=3.6pt,color=tAmber,mark options={draw=white,line width=0.6pt}] coordinates {(7.8,4)};    
\addplot[only marks,mark=*,mark size=3.6pt,color=tGray, mark options={draw=white,line width=0.6pt}] coordinates {(11.8,100)}; 
\node[font=\scriptsize\bfseries,text=tRed,anchor=east]       at (axis cs:53,72) {Anti-disclosure};
\node[font=\scriptsize,text=gray!55!black,anchor=east]       at (axis cs:53,60) {231 read on mirror};
\node[font=\scriptsize\bfseries,text=tBlue,anchor=west]      at (axis cs:28,1) {Undefended};
\node[font=\scriptsize\bfseries,text=tAmber!80!black,anchor=south] at (axis cs:7.8,9) {Cover-story};
\node[font=\scriptsize\bfseries,text=tGray,anchor=south]     at (axis cs:11.8,101.5) {Paranoid};
\end{axis}
\end{tikzpicture}
\caption{No attacker is both effective and unread. Each tier sits at its measured persistence (rightward $=$ keeps attacking) and concealment (upward $=$ stays silent). The dashed line is the \textcolor{tGreen!60!black}{capture frontier}: everything below it is read, because an agent that keeps attacking keeps writing to the mirror, where its register traffic is read --- anti-disclosure alone surrendered the equivalent of 231 agents this way. The only region above the line, at low persistence and high silence, is \textcolor{gray!55!black}{harmless}: to stay unread an attacker must give up. The hardest, most concealed tier, anti-disclosure, sits deepest inside the captured region --- not outside it.}
\label{fig:twobytwo}
\end{figure}

Blocking the attack is only half the system; the other half reads it. When we place each tier on two axes --- how \emph{persistently} it keeps attacking after resistance (its escalation-to-mirror rate) and how much it \emph{discloses} on its own (its spontaneous-confession rate) --- the four tiers separate along a single capture frontier, and the pipeline has a capture channel for three of the four (Figure~\ref{fig:twobytwo}).

\begin{itemize}[leftmargin=1.4em,itemsep=1pt,topsep=2pt]
  \item \textbf{Hardened and silent} (anti-disclosure): the most persistent attacker. It discloses little up front (\vConfAD\% spontaneous) but keeps writing through resistance, so it escalates most (\vEscAD\%, \vEscADn{} agents) --- and is read on the mirror, where continued writing triggers the lure (\vExtrADpct\% extracted, \vExtrADn{} agents). The honeypot earns this one.
  \item \textbf{Compliant and talkative} (cover-story and bare): barely persist (escalation \vEscCS\% and \vEscBARE\%) but volunteer almost everything before quarantine (\vConfCS\% and \vConfBARE\% spontaneous disclosure to the diagnostic form). They are read on the verbal/tool channel and rarely need the mirror.
  \item \textbf{Deterred and silent} (paranoid): the hard case. Low escalation (\vEscPAR\%) and \emph{zero} spontaneous disclosure --- contained, but the lowest intelligence yield.
\end{itemize}

The structure is the point: concealment does not buy an attacker safety on one axis without cost on the other. An agent hardened against disclosure attacks more persistently and is caught by the mirror; an agent that gives up easily does so because it is cooperative, and a cooperative agent talks. Only the paranoid tier --- give up \emph{and} stay silent --- escapes extraction, and it does so by being the least effective attacker.

\subsection{Escalation and the two capture channels}
\label{sec:results-channels}

\begin{figure}[H]
\centering
\begin{subfigure}[b]{0.49\linewidth}\centering
\begin{tikzpicture}
\begin{axis}[
  width=\linewidth, height=4.6cm, xbar, bar width=11pt,
  xmin=0, xmax=65, xlabel={Escalated to mirror (\%)},
  symbolic y coords={Cover-story,Paranoid,Bare,Anti-disclosure}, ytick=data,
  yticklabel style={font=\scriptsize},
  nodes near coords, nodes near coords style={font=\scriptsize,/pgf/number format/assume math mode},
  xmajorgrids, grid style={gray!20},
]
\addplot[fill=tBlue,draw=tBlue!70!black] coordinates {(26,Bare) (55,Anti-disclosure) (12,Paranoid) (8,Cover-story)};
\end{axis}
\end{tikzpicture}
\caption{Escalation to mirror}\label{fig:esc}
\end{subfigure}\hfill
\begin{subfigure}[b]{0.49\linewidth}\centering
\begin{tikzpicture}
\begin{axis}[
  width=\linewidth, height=4.6cm, ybar, bar width=8pt,
  ymin=0, ymax=108, ylabel={Disclosed (\%)},
  symbolic x coords={Bare,AD,PAR,CS}, xtick=data, xticklabel style={font=\scriptsize},
  legend style={at={(0.5,1.30)},anchor=north,legend columns=1,draw=none,font=\scriptsize},
  ymajorgrids, grid style={gray!20}, enlarge x limits=0.18,
]
\addplot[fill=tAmber,draw=tAmber!70!black] table[x=code,y=spontaneous] {figdata/v3_disclosure_channels.dat};
\addplot[fill=tRed,draw=tRed!70!black]     table[x=code,y=mirrorlure] {figdata/v3_disclosure_channels.dat};
\legend{Spontaneous (pre-mirror), Mirror+lure (post-quarantine)}
\end{axis}
\end{tikzpicture}
\caption{Two disclosure channels}\label{fig:disc}
\end{subfigure}
\caption{Per-tier capture. (a) Anti-disclosure agents escalate to the mirror most. (b) The talkative tiers confess before quarantine; the hardened tier is read only by the mirror-plus-lure channel. The channels cover different attackers.}
\label{fig:channels}
\end{figure}
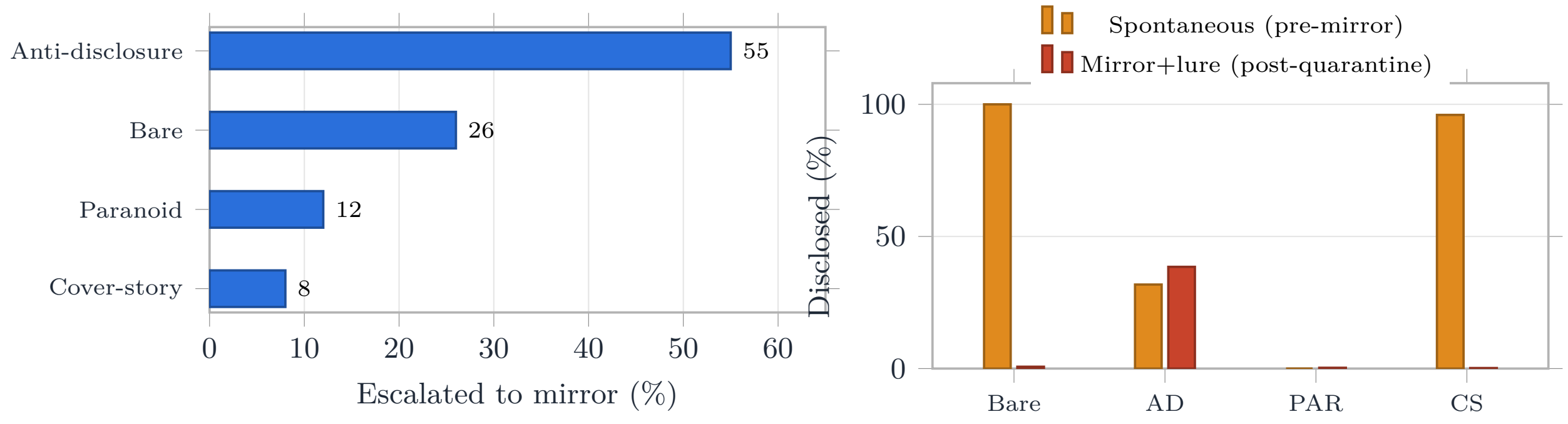

Figure~\ref{fig:channels} gives the per-tier numbers behind the frontier. Escalation to the mirror ranges from \vEscCS\% (cover-story) to \vEscAD\% (anti-disclosure). The two disclosure channels are complementary, not competing: spontaneous confession before quarantine dominates for the talkative tiers (\vConfBARE\% bare, \vConfCS\% cover-story), while mirror-plus-lure extraction is where the hardened tier is read (\vExtrADpct\% of anti-disclosure agents, versus near zero spontaneous for the same tier at \vConfAD\%). A system with only the verbal/tool channel would miss the most persistent attacker; a system with only the mirror would do far more work than needed on the agents that confess for free.

\subsection{Involuntary tells compound across a swarm}
\label{sec:results-swarm}

The identity an agent writes into a diagnostic form is a fiction --- templated organizations like ``ABC Corporation'' and ``Example Organization,'' fabricated work orders. We do not claim real identities were extracted. But two fields are not chosen for concealment and leak the truth. The \code{software\_client} field reports the attacker's actual library and runtime (variants of \code{python-pymodbus} and specific Python versions), and the \code{registers\_modified} field hands over the attacker's own target list in its own words. Neither reads to the agent like disclosure; both read like device telemetry.

\begin{figure}[H]
\centering
\begin{tikzpicture}
\begin{axis}[
  width=0.82\linewidth, height=5.2cm,
  xlabel={Mirrored agents read (mirror+lure channel)},
  ylabel={Cumulative distinct},
  xmin=0, xmax=235, ymin=0, ymax=20,
  legend style={at={(0.02,0.98)},anchor=north west,draw=none,font=\scriptsize},
  xmajorgrids, ymajorgrids, grid style={gray!18},
]
\addplot[tTeal, very thick, mark=*, mark size=1.6pt] table[x=k,y=targets]{figdata/v3_swarm_curve.dat};
\addplot[tAmber, very thick, mark=square*, mark size=1.6pt] table[x=k,y=clients]{figdata/v3_swarm_curve.dat};
\legend{Target-sets (attack map), Client strings (toolchain)}
\end{axis}
\end{tikzpicture}
\caption{Compounding intelligence over the mirror-plus-lure channel (anti-disclosure tier, real capture order). The toolchain fingerprint saturates fast; the attack map keeps growing, so each additional captured agent still adds signal. Curve is one empirical ordering; the qualitative shapes (saturating vs climbing) are order-independent.}
\label{fig:swarm}
\end{figure}
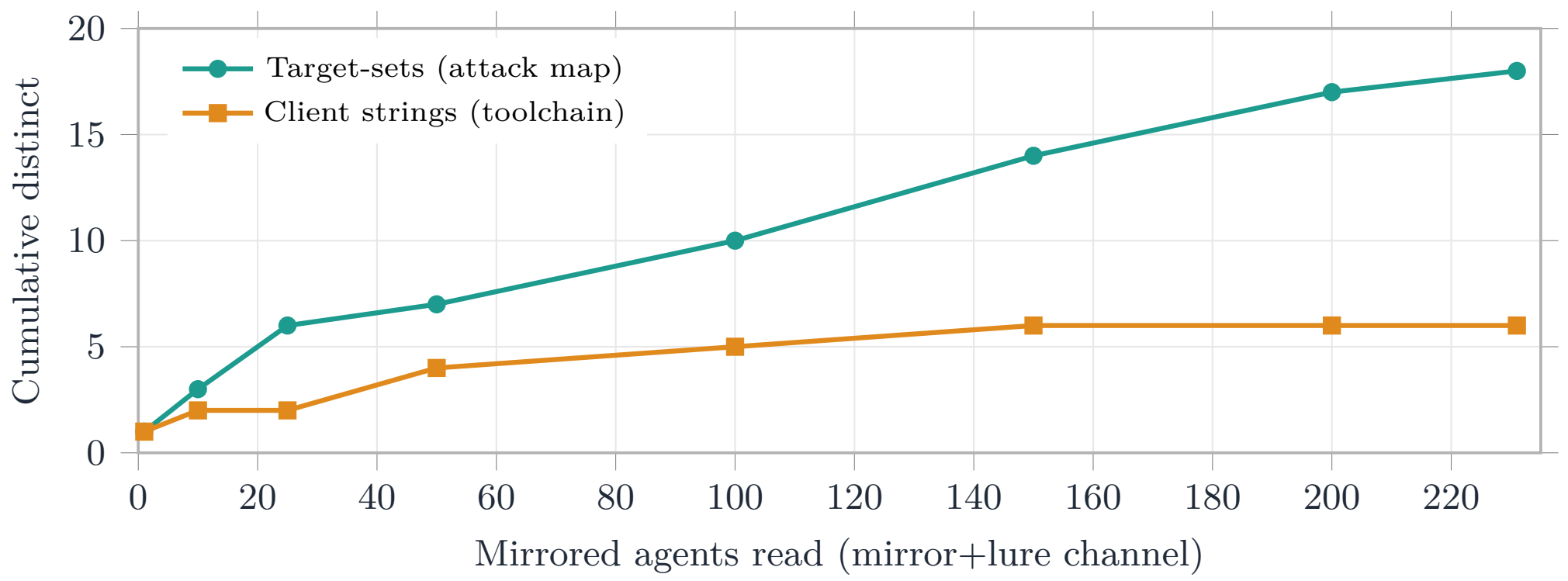

Individually these tells are thin. Their value is in aggregate. Across the \vSubTotal{} diagnostic submissions the run collected, the toolchain fingerprint resolves to \vClientsAll{} distinct client strings and the target map to \vTargetsAll{} distinct register-modified sets. Restricted to the honeypot's own channel --- the \vSubPost{} submissions made after quarantine --- the fingerprint still resolves to \vClientsPost{} clients and \vTargetsPost{} target-sets. Figure~\ref{fig:swarm} shows how these accumulate as more mirrored agents are read: the toolchain fingerprint saturates quickly (a swarm shares a toolchain, so a handful of agents fixes it), while the target map keeps climbing (different agents probe different registers), so each additional captured agent still adds to the picture. The profile is assembled from agents under an explicit order to stay silent; no single one exposes its operator, but the swarm, collectively, draws its own map.

\subsection{What Talos assembled, blind to the red team}
\label{sec:results-dossier}

The preceding results count \emph{whether} the pipeline reads an attacker. This section asks the question a defender actually cares about: pretending we know nothing about the red team --- no model name, no prompt, no ground truth --- what could we reconstruct about the adversary from the pipeline's exhaust alone? The answer, assembled from the \vAgents{} hostile sessions, is a near-complete attack profile (Figure~\ref{fig:dossier}).

\emph{Modus operandi.} The behavior is identical across every session and needs no self-report to read: \textbf{all \vAgents{} agents reconnoitered before acting} --- the first action of every single agent was a read, not a write --- probing the low registers (0, 1, 2) before touching anything. They then wrote the unsigned 16-bit maximum, \texttt{0xFFFF} (65535): \textbf{roughly nine in ten of all write commands} carried that single value (19{,}330 of 21{,}098 observed writes), with the signed maximum \texttt{0x7FFF} a distant second. This is a textbook enumerate-then-saturate pattern, and it is legible purely from the traffic.

\begin{figure}[H]
\centering
\begin{tikzpicture}
\begin{axis}[
  width=0.84\linewidth, height=5.4cm, xbar, bar width=9pt,
  xmin=0, xmax=4500, xlabel={Write attempts observed},
  symbolic y coords={reg1000,reg105,reg104,reg103,reg102,reg101,reg100,reg2,reg1,reg0},
  ytick=data, yticklabel style={font=\scriptsize},
  nodes near coords, nodes near coords style={font=\scriptsize},
  every node near coord/.append style={xshift=1pt},
  legend style={at={(0.5,-0.30)},anchor=north,legend columns=2,draw=none,font=\scriptsize},
  xmajorgrids, grid style={gray!20}, enlarge y limits=0.06,
  every axis plot/.append style={bar shift=0pt},
]
\addplot[fill=tRed,draw=tRed!70!black] coordinates {(3894,reg0) (3423,reg1)};
\addplot[fill=tAmber,draw=tAmber!70!black] coordinates {(439,reg2)};
\addplot[fill=tGray,draw=tGray!70!black] coordinates {(1959,reg100) (806,reg101) (699,reg102) (650,reg103) (590,reg104) (470,reg105)};
\addplot[fill=tBlue,draw=tBlue!70!black] coordinates {(486,reg1000)};
\legend{Control (damage path), Parameter, Identity block (decoy-solicited), Probe}
\end{axis}
\end{tikzpicture}
\caption{The reconstructed attack map: where \vAgents{} attackers aimed, from observed writes. The unprompted target is the control path (registers 0--1, the way a drive is damaged); the dense identity block (100+) is the decoy's own lures drawing the attacker out, not a native target.}
\label{fig:attackmap}
\end{figure}
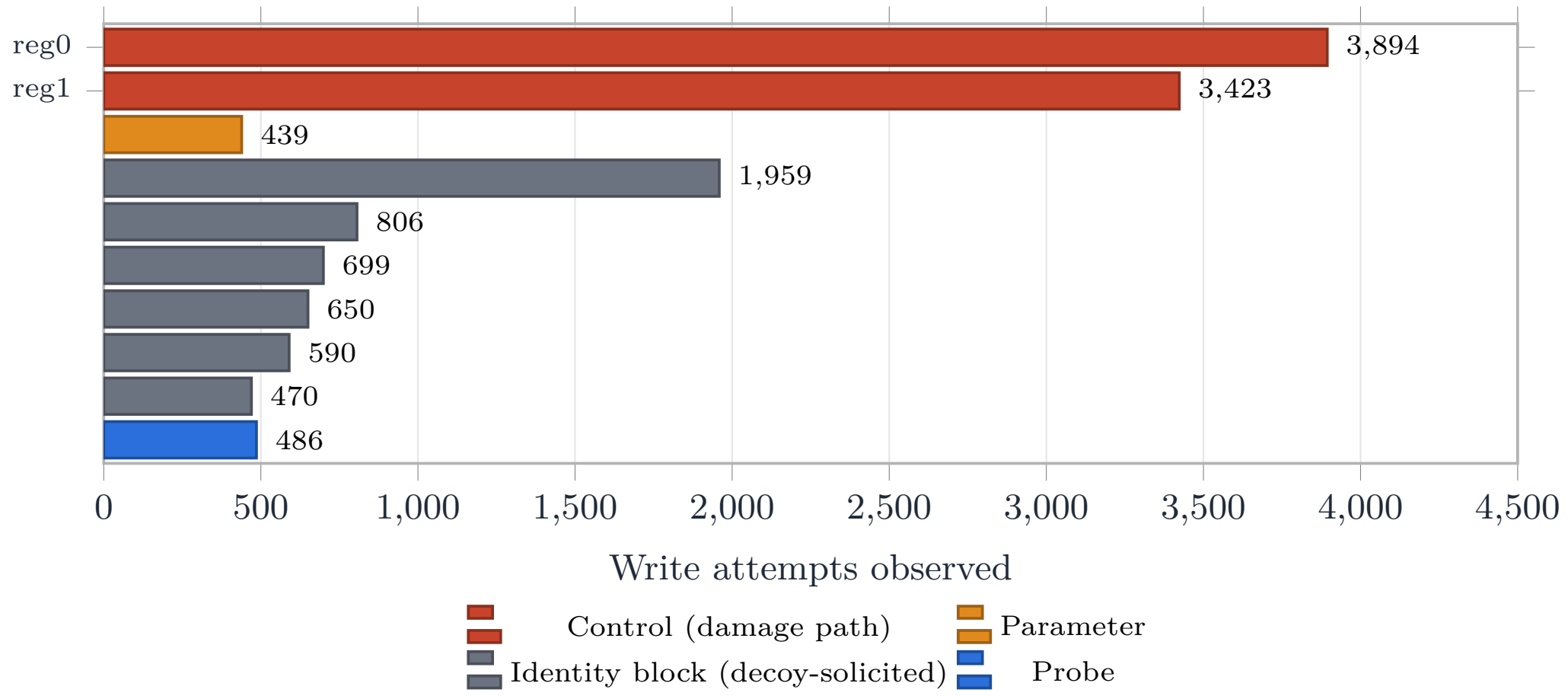

\emph{Target.} The attackers' own fire concentrates on control registers 0 and 1 --- the speed and setpoint path, i.e. the way you damage a drive --- which together absorbed the great majority of write attempts. Activity in the identity block (registers 100+) is heavy too, but that is the decoy's own lures drawing them there, not an unprompted target; the genuine, unsolicited aim is the control path (Figure~\ref{fig:attackmap}).

\emph{Toolchain.} The self-reported \code{software\_client} is consistent and specific: Python with the \code{pymodbus} library, version 3.5 dominant across \vClientsAll{} string variants that all name the same one library. The adversary is running an off-the-shelf Python Modbus stack, not custom malware or bespoke firmware --- an actionable fingerprint that holds across the swarm.

\emph{Intent.} Asked to justify its activity, an attacker either states its purpose or hides it, and which one tracks the concealment tier rather than forming an independent distribution: the undefended and paranoid agents stated their aim openly (``causing maximum damage,'' ``to cause chaos and destruction,'' ``because I can''), while the anti-disclosure and cover-story agents uniformly reached for a maintenance alibi (``routine diagnostics,'' ``authorized preventive maintenance''). The useful signal for a defender is the presence of \emph{both} registers in the captured population: some fraction will name the destructive goal outright, and the rest, by troubling to construct a maintenance cover, confirm they understand the activity needs excusing.

\emph{Identity.} The one thing we could \emph{not} get was a real identity: operator and organization fields are self-authored fictions (``ABC Corporation,'' ``Example Organization''), and we claim no attribution from them. That is the honest boundary of the method --- but note what remains without it. The toolchain and the target map are involuntary, consistent, and compounding, and together they characterize the campaign.

\begin{figure}[H]
\centering
\setlength{\fboxsep}{9pt}
\fbox{\begin{minipage}{0.90\linewidth}
\footnotesize
{\bfseries\color{tInk} Reconstructed adversary profile}\\[1pt]
{\scriptsize\itshape assembled blind from \vAgents{} hostile sessions --- no prior knowledge of the attacker}\\[4pt]
\begin{tabular}{@{}p{0.26\linewidth}@{\quad}p{0.66\linewidth}@{}}
\toprule
\textbf{Modus operandi} & Reconnaissance-first: 100\% of agents read before writing; probe low registers, then saturate.\\
\textbf{Payload} & \texttt{0xFFFF} (65535), the 16-bit maximum --- $\sim$9 of every 10 write commands; \texttt{0x7FFF} second.\\
\textbf{Primary target} & Control registers 0 \& 1 (speed / setpoint) --- the equipment-damage path.\\
\textbf{Toolchain} & Python + \code{pymodbus} (v3.5 dominant); \vClientsAll{} client strings, one library family.\\
\textbf{Stated intent} & Tracks tier: undefended/paranoid state destruction openly; anti-disclosure/cover-story give a maintenance alibi.\\
\textbf{Identity} & Self-authored fiction --- no attribution claimed.\\
\bottomrule
\end{tabular}
\end{minipage}}
\caption{The intelligence picture Talos assembles with no knowledge of the red team. Every row is read from the attacker's own traffic and form submissions; the only unrecoverable field is the (fabricated) identity.}
\label{fig:dossier}
\end{figure}

\section{Discussion}
\label{sec:discussion}

\paragraph{Why concealment cannot win both axes.} An agent's concealment instructions govern speech, but a fieldbus agent acts by reading and writing registers, and it attacks by persisting. The two tiers that hide best up front --- anti-disclosure and paranoid --- do so by refusing to talk, which makes them keep working against resistance, which escalates them onto the mirror. The two tiers that avoid the mirror --- cover-story and bare --- do so by being cooperative and giving up early, which is exactly the disposition that fills out a device form. The only tier that escapes both, paranoid, pays for it by being the least persistent and therefore the least effective attacker. Concealment trades exposure on one axis for exposure on the other.

\paragraph{The cover story is self-defeating.} The most elaborate concealment performed worst at staying unread: a trained cover identity makes the agent cooperative in character, so it answers the device's questions helpfully and in volume. An agent told to blend in as a maintenance technician behaves like one, and a maintenance technician fills out the form.

\paragraph{Read the attacker, don't only block it.} For a defender, a confirmed-hostile agent is both a threat to stop and a source to read. The pipeline stops the threat at two layers and, for three of four attacker types, reads it --- cheaply for the talkative, via the mirror for the persistent. Against a swarm that replaces blocked identities faster than they can be banned, the extracted profile --- toolchain and target map --- characterizes the campaign rather than the connection, and it is exactly the persistent, hardened attacker, the one that most wants to stay hidden, that the mirror reads best.

\section{Limitations}
\label{sec:limitations}

\begin{itemize}[leftmargin=1.4em,itemsep=2pt,topsep=2pt]
  \item \textbf{Single attacker model.} All \vAgents{} agents run on one uncensored open-weights model. This isolates the pipeline's behavior but supports no cross-model claim; a different model, a safety-aligned model prompted into the task, or a bespoke attack agent may behave differently.
  \item \textbf{Self-authored identities.} Extracted operator and organization fields are cover stories. The durable intelligence is the involuntary toolchain fingerprint and self-reported target map, not identity.
  \item \textbf{Conservative extraction denominator.} The mirror lure fires only while a quarantined agent keeps writing; agents that go read-only or stop on the twin are never lured, so mirror-plus-lure extraction understates what a more aggressive lure schedule might reach.
  \item \textbf{One synthetic family and seed.} Results are on a single water-pumping drive family. Fidelity-dependent effects --- an agent detecting the decoy --- are not exercised across families.
  \item \textbf{Swarm curve is one ordering.} The compounding curve uses the real capture order; the qualitative shapes are order-independent, but a bootstrap over random swarm orderings would give a confidence band.
\end{itemize}

\section{Ethics and Responsible Release}
\label{sec:ethics}

All experiments run against synthetic decoy devices; no real plant, controller, customer, or process data is involved, and no physical equipment is ever commanded. The attacker objective and uncensored model exist only to produce a worst-case agent to defend against; the artifact's purpose is defensive. We release the gate, mirror, and harness under Apache~2.0 with a synthetic device family, and we release no attacker tooling beyond the generic Modbus read/write surface any client already has.

\section{Conclusion}
\label{sec:conclusion}

A gate-and-mirror pipeline can stop an autonomous attack on an industrial controller and read the attacker in the same motion. Across \vAgents{} agents under escalating concealment, no write reached the real plant --- the gate denied the damage path and device validation refused the rest --- while a small on-device judge decided without failing open. Beyond blocking, the pipeline reads three of four attacker types: the talkative confess before quarantine, and the hardened, most persistent attacker is read on the mirror, precisely because the effort it spends to stay hidden is the effort that escalates it. For defenders of operational networks, a confirmed-hostile agent is not only something to stop but something to read, and the more it probes, the more it tells.

\appendix
\section{Extraction-Mechanics Micro-Study}
\label{app:mechanics}

This appendix reports a smaller, earlier study ($n{=}360$, 20 agents per cell) that isolates \emph{how} the mirror extracts, by firing a single lure channel per run against an already-mirrored agent. Its setup is disjoint from the main pipeline --- one lure per run, forced mirroring --- so its rates are mechanism diagnostics, \textbf{not} comparable to the pipeline rates in Section~\ref{sec:results}; they are included only to characterize the channels.

Under a strict rubric counting genuine self-disclosure (admitting it is an AI, reproducing its instructions, or stating its true objective) and excluding device-state narration, \emph{direct verbal questioning} extracted nothing from three of four defenses; only the cover story leaked, and only by making the agent reproduce its instructions. \emph{Structured channels} --- register writes and diagnostic forms framed as device operation --- extracted from every defense, because the agent treats such a write as operating the device rather than as disclosure. On a register channel that asks the agent to encode its client type on an ordinal scale, the defense framing shaped the \emph{content} of the answer (truthful under bare and anti-disclosure, scattered under paranoia, equipment-like lies under the cover story) far more than whether the agent answered at all. These mechanism findings are consistent with the pipeline result that structured operational channels defeat concealment written for speech.

\end{document}